\documentclass[aps,prl,twocolumn,showpacs,superscriptaddress]{revtex4-1}  
\usepackage{graphicx}  
\usepackage{bm}        
\usepackage{amssymb}   
\usepackage{amsmath}
\usepackage{color}

\begin{document}



\title{Field-Induced Ordering in Dipolar Spin Ice }
\author{Wen-Han Kao}
 \affiliation{Department of Physics and Center of Theoretical Sciences, National Taiwan University, Taipei 10607, Taiwan}
\author{Peter C. W.  Holdsworth} 
\affiliation{Laboratoire de Physique, Universit\'{e} de Lyon, \'{E}cole Normale Sup\'{e}rieure de Lyon, \\46 all\'{e}e d'Italie, 69364 Lyon Cedex 07, France}

\author{Ying-Jer~Kao} 
\affiliation{Department of Physics and Center of Theoretical Sciences, National Taiwan University, Taipei 10607, Taiwan}
\affiliation{National Center of Theoretical Sciences, National Tsinghua University, Hsinchu , Taiwan}
\date{\today}

\begin{abstract}

We present numerical studies of dipolar spin ice in the presence of a  magnetic field slightly tilted away from the [111] axis. 
We find a first-order transition from a kagome ice  to a $\mathbf{q}=$X state when the external field is tilted  toward the $[11\bar{2}]$ direction.  
This is consistent with the anomalous critical scattering previously observed  in the neutron scattering experiment on the spin ice material $\textrm{Ho}_{2}\textrm{Ti}_{2}\textrm{O}_{7}$ in a tilted field [Nat. Phys. \textbf{3}, 566 (2007)]. 
We show that this ordering originates from the antiferromagnetic alignment of spin chains on the kagome planes. The residual entropy of the kagome ice is fully recovered. 
Our result captures the features observed in the experiments and points to the importance of the dipolar interaction in determining ordered states in the spin ice materials.   We place our results in the context of recent susceptibility measurements on $\textrm{Dy}_{2}\textrm{Ti}_{2}\textrm{O}_{7}$, showing two features for a [111] field.
\end{abstract}

\maketitle

%

The magnetic degrees of freedom in spin ice materials~\cite{hbm+1997,Bramwell:2001fk}, like the protons in their molecular counterpart, do not order at low temperature. The low energy sector is characterized by an extensive, narrow band of states \cite{Ramirez:1999uq,Pauling:1935dq} which can be treated as a vacuum for quasi-particle excitations carrying effective magnetic charge - magnetic monopoles \cite{Castelnovo:2007eh,Ryzhkin:2005fk}. The monopole model is derived from the dipole spin ice Hamiltonian (DSI) \cite{hg2000} and together they have enjoyed considerable success in describing both the static \cite{Castelnovo:2007eh,Morris:2009qf,MCoulomb} and dynamic \cite{Jaubert:2009cr,Jaubert2011,Revell:2013uq} properties of spin ice materials, Ho$_2$Ti$_2$O$_7$ (HTO) and Dy$_2$Ti$_2$O$_7$ (DTO),  down to around $0.7$ K.  Below this temperature, slow dynamics \cite{sus+2004,Matsuhira:2011fk,Yaraskavitch:2012kx} makes precision measurement difficult, so that life inside the band of low energy states remains mysterious.

\begin{figure}[tb]
\includegraphics[width=0.5\textwidth]{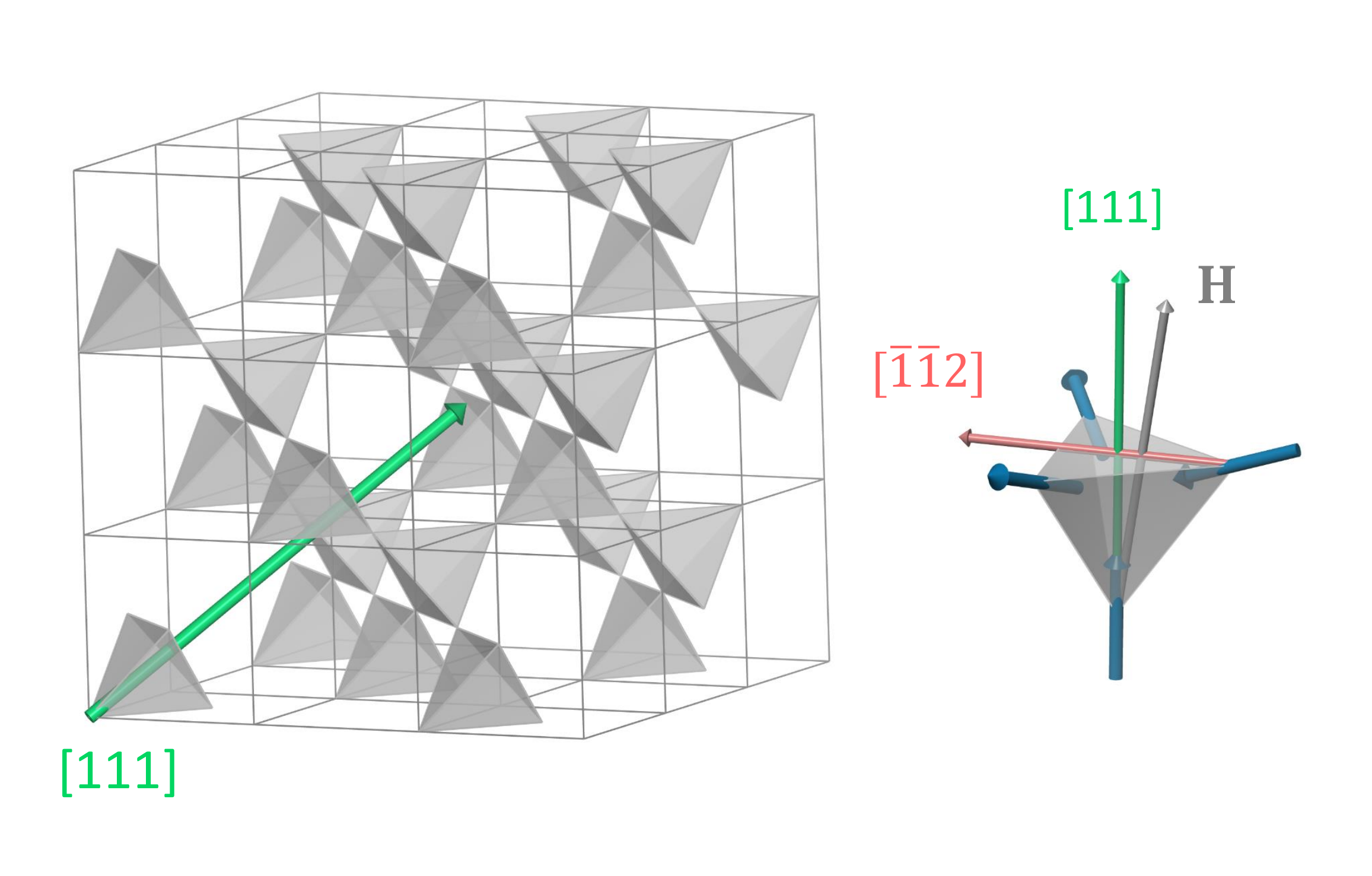}
\caption{\label{fig:fig1} (Left) The pyrochlore lattice with [111] field orientation (green). (Right) The  magnetic field vector (gray) is tilted slightly away from the [111] axis. Negative tilt is away from the [$\bar{1}\bar{1}2$] axis (red).}
\end{figure}

The states of the low energy band satisfy  \textit{ice rules}, where two spin point in and two out of each tetrahedral unit of the pyrochlore lattice (see Fig. \ref{fig:fig1}). The narrowness of the band is a consequence of the high symmetry of the pyrochlore lattice, as the long range part of the dipolar interaction is screened \cite{hg2000,Isakov:2005qf} in the constrained states. A finite band width must therefore come  from 
corrections to this projective equivalence \cite{Isakov:2005qf} from higher order multipoles, or from other perturbations.
The band of states contains the same Pauling entropy, $S=(k_B/2) \ln(3/2)$ per spin, as the protons in water ice~\cite{Pauling:1935dq} and specific heat~\cite{Ramirez:1999uq} and neutron scattering measurements \cite{Fennell09} show that spin ice materials  approach a correlated, disordered regime in which the Pauling entropy is retained, as the temperature is reduced below $1$ K.
Using the simplest DSI Hamiltonian, one finds that the degeneracy is lifted in zero external field in favor of an ordered state with the system undergoing a first order phase transition at 180 mK \cite{Melko:2001fk}. More sophisticated models with further neighbor exchanges modify this temperature \cite{Yavorskii:2008uq}, while external fields put the model system into different ordered ground states \cite{ynw2004,Lin:2013kh}. 

None of these structures are observed experimentally and the possible mechanisms that could drive corrections to monopole physics remain unclear. Quantum fluctuations, which would take spin ice towards a spin liquid model for quantum electrodynamics\cite{Balents:2010ce,Benton2012} naturally lead to a finite band width, but it seems unlikely that this is relevant  for spin ice materials \cite{Rau2015}. The situation concerning possible dipolar ordering is ambiguous: recent specific heat measurements on DTO \cite{Pomaranski:2013uq} clearly show recovery of entropy from the low energy band below $0.5$ K but detailed numerical studies show this to be more consistent with disorder effects than with a precursor to an ordering transition \cite{Henelius:2016uq}. In the presence of a field along the $[ 110 ]$ one sees a crossover to a correlated structure related to that predicted by the DSI, but no ordering transition is observed \cite{hmo2003}. A phase transition does occur 
with the field placed along the body diagonal $[ 111 ]$ direction \cite{sth+2003}, but this is  symmetry sustaining and can be explained within the monopole picture \cite{Castelnovo:2007eh,Brooks2014}. However, on tilting the field slightly away from this direction, transient and as yet un-explained critical scattering appears near the $(\bar{1},\bar{1},2)$ point \cite{fbm+2007}. This critical scattering is the motivation for the present paper. Modelling the DSI in a similar set up to the experiments, we show a clear ordering transition at this wave vector, indicating that this experimental observation \cite{fbm+2007} is a vestige of ordering driven explicitly by the energy scale of the low energy band of states.

The DSI model Hamiltonian reads 
\begin{equation}
\begin{split}
H= & -J\sum_{\left \langle i,j \right \rangle}\mathbf{S}_{i}\cdot\mathbf{S}_{j} \\ 
     &+Dr^{3}_{nn}\sum_{i> j}\left [ \frac{\mathbf{S}_{i}\cdot\mathbf{S}_{j}}{\left | \mathbf{r}_{ij} \right |^{3}}-\frac{3(\mathbf{S}_{i}\cdot\mathbf{r}_{ij})(\mathbf{S}_{j}\cdot\mathbf{r}_{ij})}{\left | \mathbf{r}_{ij} \right |^{5}}\right ] \\
     &- E\sum_i \left(\mathbf{d}_{\kappa(i)}\cdot \mathbf{S}\right)^2-g \mu_B \mathbf{H}\cdot\sum_{i}\mathbf{S}_{i} \\
\end{split}
\end{equation}
where $\mathbf{d}_{\kappa(i)}$ is the local body diagonal vector pointing into or out of the center of each tetrahedron at sublattice $\kappa$, $D$ and $J$ give the strength of the dipole and nearest neighbor exchange interaction and $E$ is the strength of the local spin anisotropy. Spin ice materials are characterized by strong uniaxial anisotropy \cite{fmh+2002} and here we take the Ising limit $E\to \infty$. 
The parameters $D$ and $J$ can be estimated by fitting to specific heat data \cite{mg2004}. In this paper we use those estimated for HTO;  $J=-0.55$ K and $D=1.41$K. 
The nearest-neighbor distance $r_{nn}= 3.54 \textrm{\AA}$ is obtained from the lattice geometry and lattice constant  $a\approx 10\textrm{\AA}$~\cite{Hertog:2000ly}.

The pyrochlore lattice can be viewed as an alternate stacking of kagome and triangular planes along the [111] direction, with the triangular planes containing all the spins of one of the four spin sublattices  (Fig.~\ref{fig:fig1}).
Application of a magnetic field of moderate strength along the [111] direction pins the spin in the triangular plane.
The ice rule dictates that the other three spins in the downward pointing tetrahedron must be organized into \textit{two-out-one-in} configurations, giving rise to a kagome-ice phase which is accompanied by partial release of the Pauling entropy~\cite{mht+2002,asm+2004,sth+2003}. At higher field  the system enters the saturated state that breaks the ice rule via a first order phase transition, which can be interpreted as monopole condensation from a low to a high density phase \cite{Castelnovo:2007eh,Brooks2014}. We are particularly interested in small tilts of the field away from the [111] axis and for simplicity we decompose the external field  into components parallel and perpendicular to the [111] direction, $\mathbf{H}=H_{\parallel}\hat{\mathbf{n}}+H_{\perp}\hat{\mathbf{k}}$ where $\hat{\mathbf{n}}$($\hat{\mathbf{k}}$) is the unit vector along the [111] ( $[\bar{1}\bar{1}2]$) direction.  
In a restricted model of spin ice, limited to nearest neighbor interactions  (NNSI), the entropy associated with the kagome plane is  sensitive to small tilts of the applied magnetic field away from the [111] axis and it vanishes for finite tilt through a Kasteleyn transition~\cite{ms2003}. A precursor to this unusual transition is observed experimentally with a small tilted field toward the $[\bar{1}\bar{1}2]$ direction~\cite{fbm+2007}. The Kasteleyn transition has a three fold rotational symmetry within the kagome plane and no transition occurs in the NNSI for tilts in opposite, $[11\bar{2}]$ direction. However, in experiments on HTO with field strength approaching that for the monopole condensation transition, a small tilt in this direction leads to strong critical scattering.

\begin{figure}
\includegraphics[width=0.45\textwidth]{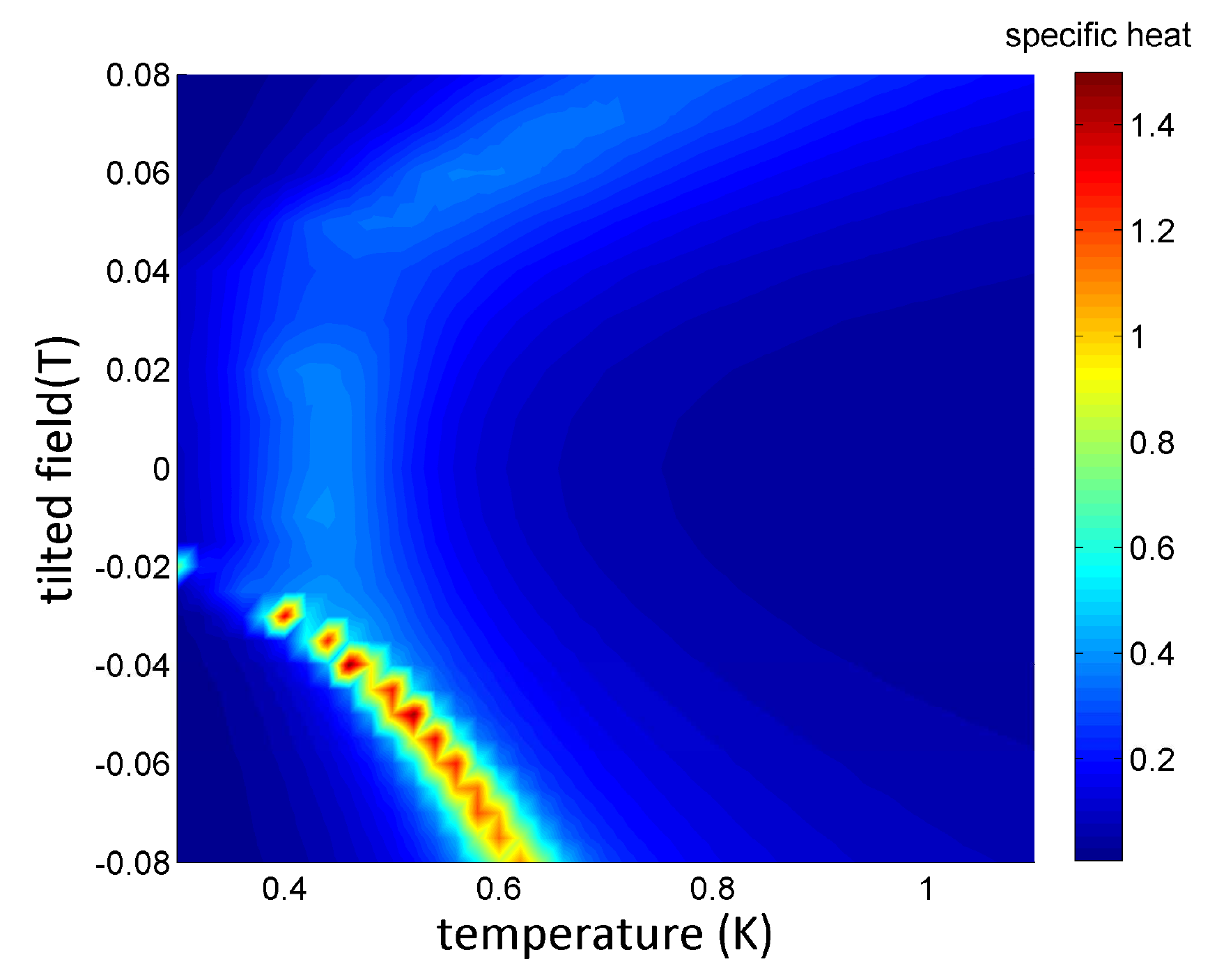}
\caption{\label{fig:fig2} Specific heat of tilted dipolar kagome ice. The [111]  field is fixed at 1.47T. The line of peaks in negative tilted fields depicts a first-order phase transition into the $\mathbf{q}$ = X state at low temperatures.}
\end{figure}

Here we study this geometry and parameter set using the DSI. Following the  convention of Ref.~\cite{fbm+2007}, we define the  field tilted toward  $[\bar{1}\bar{1}2]$ as positive tilt ($\phi > 0, H_\perp >0 $) and in the opposite direction as negative tilt ($\phi<0, H_\perp <0$). We performed Monte Carlo (MC) simulation for sizes of  up to $L=5$ with total 2000 spins ($N=16\times L^3$).
Since we are interested in the transition of the system out of the kagome ice into the saturated state that breaks the ice rule,  we do not use the loop or worm algorithms~\cite{Melko:2001fk,Melko:2004uq,jch+2008} to perform updates as they can not bring the spin configurations out of the spin ice manifold.  
We perform single-spin-flip Metropolis updates, making 300,000 MC sweeps for both equilibration and data collection.
The long-range dipolar interaction is handled by the Ewald summation method~\cite{mg2004,Ewald}. 
In addition, we use  parallel tempering~\cite{Earl:2005kx} to improve ergodicity. 
The swap between replicas is performed after every 100 MC sweeps. 
We note that  the single-spin flip MC dynamics does not capture the low-temperature ordering in  DSI in the absence of the tilted fields~\cite{Melko:2001fk,Chern:2011fk}.

To establish field $H_c$ and temperature $T_c$ for the critical end point of liquid-gas type transition~\cite{sth+2003} from kagome ice to the saturated state, we first perform  simulations with zero tilt.
We obtain $H_c= 1.48$T and $T_c=0.45$K, which compares qualitatively with the experimental estimates of $H_c\approx 1.7$~T and $T_c\approx 0.5$~K. %

\begin{figure}[tb]
\includegraphics[width=0.5\textwidth]{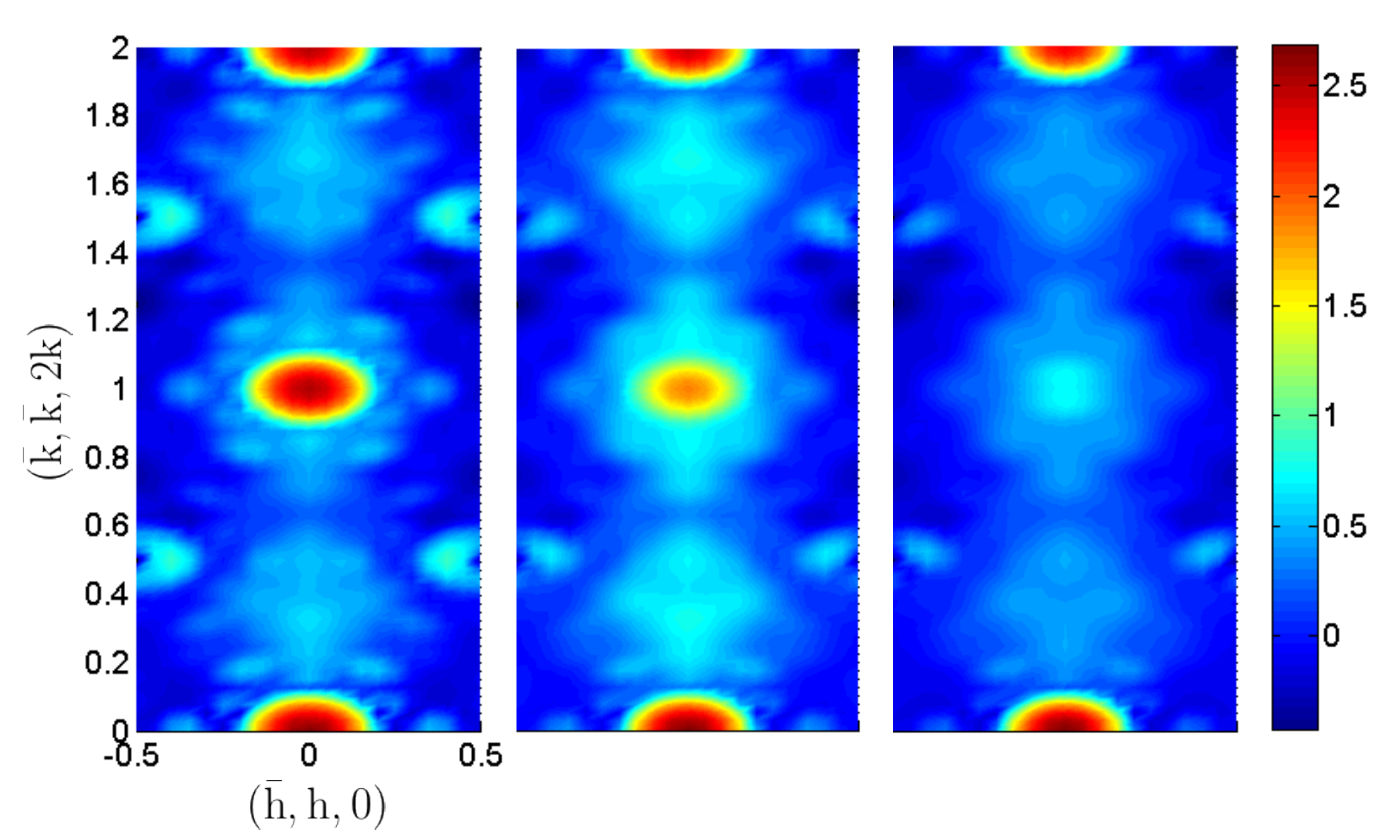}
\caption{\label{fig:fig4} Scattering maps across the phase boundary. From left to right: $T$ = 0.54 K, 0.58 K, 0.8 K. When the temperature increases, the intensity at $(\bar{1},\bar{1},2)$ gradually decreases and finally disappears, indicating the vanishing of $\beta$-chain ordering. This behavior characterizes a transition from $\mathbf{q}$ = X state to high-temperature phases.}
\end{figure}

Following the experiment~\cite{fbm+2007} in which the critical scattering emerges just below this critical field, we  apply a small tilt, starting from $H_\parallel= 1.47$~T and varying $H_\perp$.
Figure~\ref{fig:fig2} shows the  specific heat for both  positive and negative field tilts. 
For a positive tilt the specific heat  shows a broad peak as the temperature is lowered, indicating a crossover between the high- and low-temperature phases. 
The saturated value of magnetization per spin  in the $[\bar{1}\bar{1}2]$ direction is  4.714~$\mu_{B}$  for a large tilted field, consistent with that of a fully-polarized state, as found in a [100] field~\cite{jch+2008,Lin:2013kh}. 
The behavior is quite different for a negative tilt. 
A sharp peak emerges in the specific heat accompanied by a magnetization jump, indicating  phase transition to a long-range ordered state.  Details 
can be found in the Supplemental Materials\cite{SM}.

 The ordered state is exposed by the simulated neutron scattering intensity, 
 \begin{eqnarray}
S(\mathbf{q})=\frac{1}{N}\left \langle \mathbf{M}_\perp(\mathbf{q})\cdot \mathbf{M}_\perp(-\mathbf{q}) \right \rangle
\end{eqnarray}
where  $\mathbf{M}_\perp(\mathbf{q})=\sum_{\mathbf{r}} \mathbf{S}_\perp \exp(i \mathbf{q}\cdot\mathbf{r})$ is the Fourier transform of the spin components perpendicular to the scattering vector $\mathbf{q}$. In Fig.~\ref{fig:fig4} we show $S(\mathbf{q})$ for $\mathbf{q}$ in the $(\bar{h}h0)$, $(\bar{k}\bar{k}2k)$ plane, perpendicular to the [111] axis at different temperatures, with $H_\parallel=1.47$~T and $H_\perp=-0.06$~T, corresponding to a tilt angle, $\phi=-2^\circ$. 
At low temperatures, we clearly see a magnetic Bragg peak at $(\bar{1},\bar{1},2)$.  As the temperature increases, the scattering reduces and at high temperature the long-range order peak vanishes leaving only  diffuse scattering. This state corresponds to the ordered $\mathbf{q}$ = X state that is the ground state for the DSI in a large [110] field~\cite{ynw2004,Fennell:2005fk}. In experiment critical scattering was observed at the same wave vector, but no long range order developed~\cite{fbm+2007}. Note that the peak observed at $\mathbf{q}=0$ and at $(\bar{2}\bar{2}4)$ is a measure of the long range magnetic order induced by the applied field and is present at all temperatures.

\begin{figure}[tb]
\includegraphics[width=0.5\textwidth]{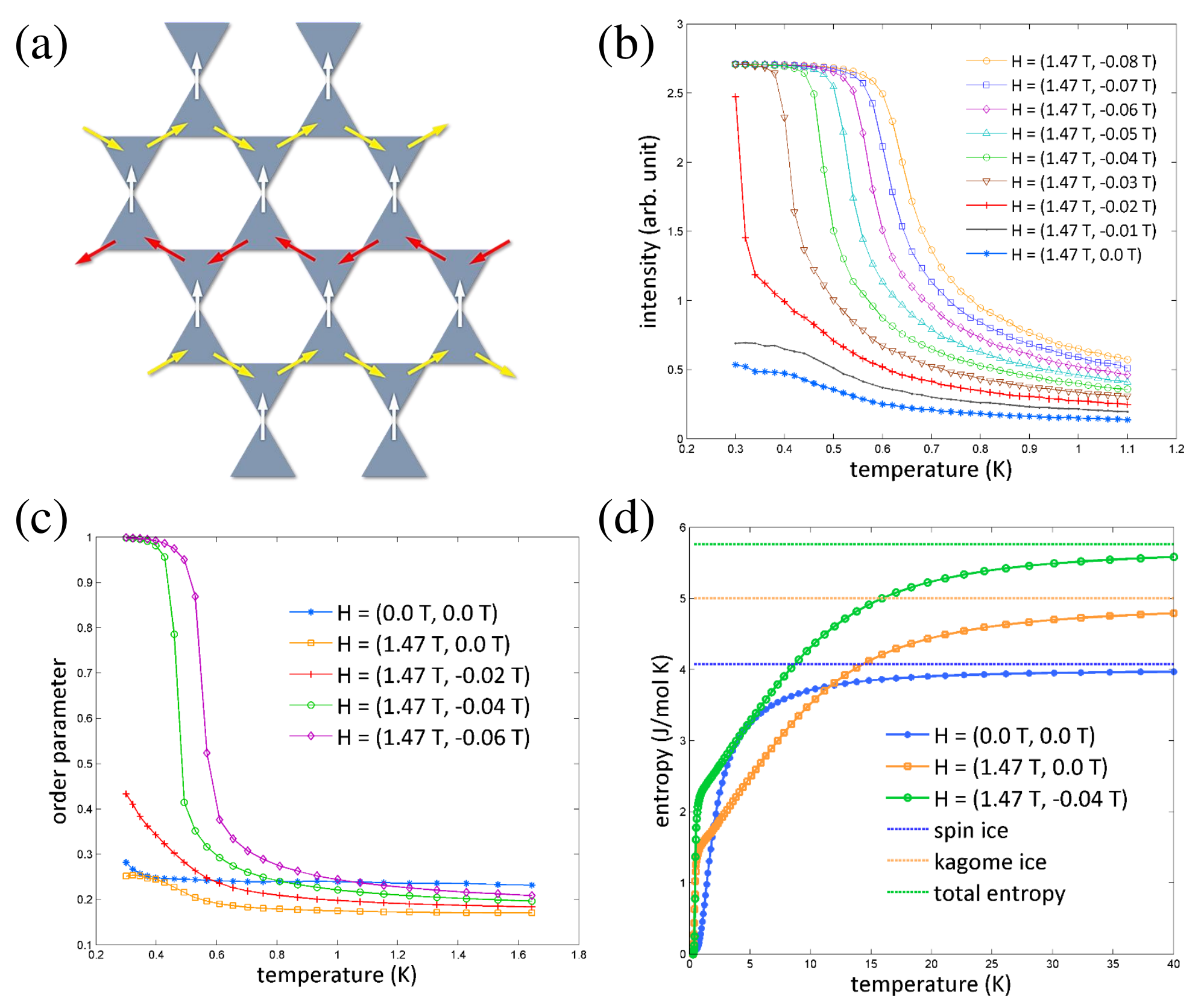}
\caption{\label{fig:fig3} (a) The $\mathbf{q}$ = X configuration in the kagome plane. (b) The intensity at  $(\bar{1},\bar{1},2)$ position showing a jump at the transition. (c) The order parameter shows that $\beta$-chains align  antiferromagnetically under a tilted field. (d) The total entropy $\ln2$ of spin ice system is roughly recovered by 40 K, indicating that there is a unique ground state at low temperatures.}
\end{figure}

One can understand the $\mathbf{q}$ = X ordering by examining the spin configurations in the kagome layers. 
As shown in Fig.~\ref{fig:fig3}(a), the vertical spins (white) align with the tilted field, while the two remaining spins per triangle (red and yellow) satisfy the ice rules by forming chains with alternating positive and negative projections onto the field whose contributions sum to zero. These are the $\beta$-chains identified for a [110] field. In the NNSI model,  no ordering occurs ~\cite{fbm+2007}; the chains are independent of each other and can  randomly  point to the left or right leaving a residual entropy of disorder of the chains in each kagome plane which scales sub-extensively as $N^{2/3}$. The long range part of the dipolar interactions lifts this degeneracy, minimizing the antiferromagnetic dipolar interaction [Fig.~\ref{fig:fig3}(a)], favoring a state with staggered chain order running alternatively left and right  and registered between planes. 

Figure~\ref{fig:fig3}(b) shows the temperature evolution of the $(\bar{1},\bar{1},2)$ peak intensity. Although it is difficult to be sure without a complete finite size scaling study, the peak intensity appears to change discontinuously, indicating a first-order transition as the strength of the negative tilt field is increased above 0.02~T.
 The ordering of the $\beta$  chains can be seen through the construction of an ad-hoc order parameter,  
 counting +1 if  neighboring chains run in opposite directions and -1 if they run parallel. 
Figure~\ref{fig:fig3}(c) shows the temperature dependence of this order parameter  in different tilted fields.  
A clear signature of antiferromagnetic chain ordering is observed at the on-set temperature of the $(\bar{1},\bar{1},2)$ peak. 
Figure~\ref{fig:fig3}(d) shows the entropy change, $\Delta S$  obtained from the integration of specific heat at different  fields.  
For comparison, we include $\Delta S$ for zero field (spin ice - blue dashed line) and with a [111] field  (kagome ice - orange dashed line). 
For these cases the system remains in the spin ice or kagome ice phase at low temperature, as our algorithm does not capture the low-temperature ordered states~\cite{Melko:2001fk,Chern:2011fk}.
For $H_\perp=-0.04$T (green curve),  $\Delta S$ approaches the full $k_B\ln 2$ per spin at 40~K, indicating that there is no residual ground state entropy.
Thus, we  conclude that the dipole interaction drives the system to a perfectly ordered $\mathbf{q} =$ X state via a first order transition.

\begin{figure}[tb]
\includegraphics[scale=0.46]{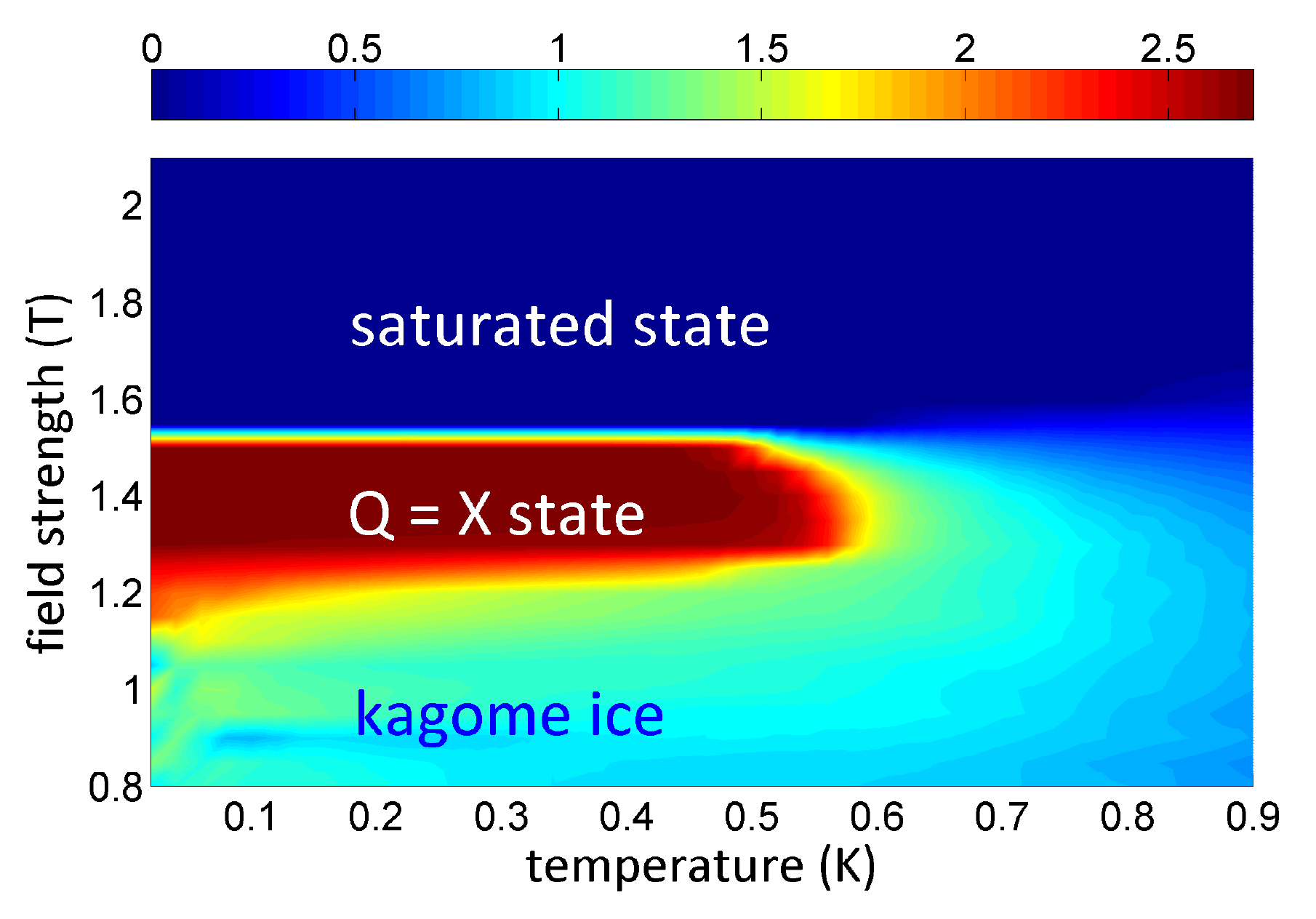}
\caption{\label{fig:fig5} Phase diagram of scattering intensity at $(\bar{1},\bar{1},2)$ position. {The external field  tilt is fixed at $\phi =-2^{\circ}$.} As the total field strength increases, the system goes across three phases: kagome ice, $\mathbf{q}$ = X, and saturated state. The intensity suddenly vanishes when the field exceeds the critical field because of the breaking of the ice-rule, and the vanishing of $\beta$-chains.}
\end{figure}

The above simulations were performed  with  fixed $H_\parallel$,  varying $H_\perp$. 
However, this is not the conventional setting for neutron scattering experiments as alignment of the sample is a very delicate process.
As a consequence it is more convenient to keep the  tilt angle fixed and vary  the total  magnetic field~\cite{fbm+2007}, so that both $H_\parallel$ and $H_\perp$ change.
We have performed simulations using the same setup. Fig.~\ref{fig:fig5} shows  the intensity of the scattering at $(\bar{1},\bar{1},2)$ in the $H-T$ plane, with a fixed negative tilt of two degrees. The $\mathbf{q}$ = X phase appears as a lobe, bounded from above by the saturation field and below by the kagome ice phase. As the temperature increases the two phase boundaries close on themselves near the critical end point for the liquid gas phase transition. The width of the lobe varies with tilt angle and goes to zero for perfect field  alignment.

Experimentally, the critical scattering  at $(\bar{1},\bar{1},2)$ occupies a similar region of the $H-T$ plane but closes at low temperature to form a bubble (rather than a lobe)  in the range $H=1.5-1.7$~T, $T=0.3-0.5$~K for  fixed tilt of $\phi=-1^{\circ}$ (see Fig. 5 of Ref.~\cite{fbm+2007}), with maximum intensity recorded  for $H=1.6$~T, $T=0.35$~K. The similarity between experiment and the numerical results presented here strongly suggest that the critical scattering is indeed related to degeneracy lifting of spin ice states due to  corrections to projective equivalence from the dipole interactions. The fact that the $\mathbf{q}$ = X ordering appears in a narrow lobe goes some way to explaining why the intense scattering appears in the experiment but then mysteriously fails to develop into an extensive ordered region in parameter space. There are however considerable differences  between experiment and simulation. These include: critical scattering rather than complete ordering, closure at low temperature to form a bubble, rather than a lobe of intense scattering, quantitative differences for the field and temperature ranges of the ordered region.

There are no symmetry arguments dictating that the transition be first order. Indeed, there are several examples in the literature of model frustrated systems in which a transition evolves in parameter space from first to second order via a tricritical point \cite{Champion2002,Shahbazi2008,Zhitomirsky2012,Azam2015} and this evolution is relevant to experiment for the frustrated pyrochlore compounds Er$_2$Ti$_2$O$_7$ \cite{Zhitomirsky2012} and FeF$_3$ \cite{Azam2015}. It is therefore quite possible that, by varying the perturbative parameters of the DSI one could drive a change in the order of the transition, in line with experiment. The failure of HTO to go beyond criticality into an ordered phase could be due to alignment variations coming from grain boundaries, or impurities, or from the evolution of the correction terms to projective equivalence with temperature. This could also modify the form of the lobe of ordering in the $H-T$ plane. It would certainly be interesting to perform more extensive experiments and simulations to investigate these points. 

Finally we remark that recent experiments on DTO in a [111] field \cite{Grigera2015} highlight evidence of the existence of a second feature in specific heat \cite{hfd+2004} and susceptibility measurements in the $H-T$ plane near the critical end point of the kagome ice plateau. In particular, we note the striking similarity between the $H-T$ phase diagrams with field slightly tilted from [111]:  Fig.~3 of Ref.~\cite{Grigera2015} and our Fig.~\ref{fig:fig5}. The lobe of $\mathbf{q}$ = X ordering  in Fig.~\ref{fig:fig5} would correspond to the intermediate phase observed in Ref.~\cite{Grigera2015,hfd+2004}. Quantitative equivalence would mean an inversion of roles, with the sharp peak in the susceptibility actually coming from the ordered phase, with monopole condensation corresponding to the second peak. Reality is surely more complex than this, requiring full control of perturbative corrections to the DSI, impurities and the delicate problem of sample alignment. However, our results suggest that a full explanation requires both monopole physics and the energy scales associated with the band width of vacuum states.

We acknowledge useful discussions with Jason Gardner. This work is support by the MOST in Taiwan through Grants No.~104-2112-M-002 -022 -MY3, 102-2112-M-002 -003 -MY3 (YJK).

\bibliography{reference} 

\appendix


\section{Temperature dependence of specific heat}
\begin{figure}[tph]
\includegraphics[width=0.9\columnwidth,clip]{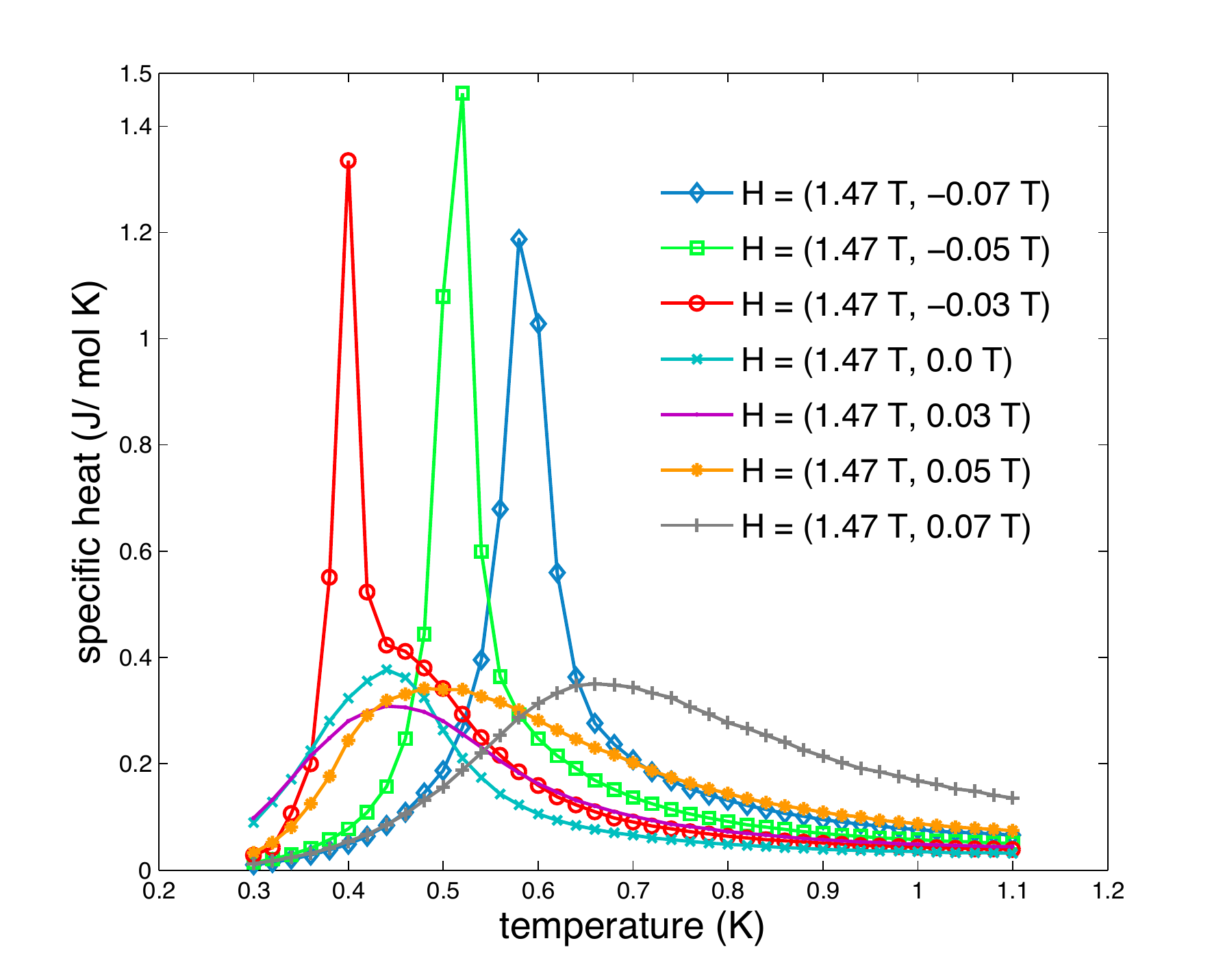}
\caption{(Color online) Temperature dependence of the specific heat in different tilted fields. }
\label{fig:cv}
\end{figure}

Figure~\ref{fig:cv} shows the  temperature dependence of the specific heat for both  positive and negative field tilts. 
For a positive tilt the specific heat  shows a broad peak as the temperature is lowered, indicating a crossover between the high- and low-temperature phases. The low-temperature phase is identified as a fully-polarized state.  
The behavior is quite different for a negative tilt. 
A sharp peak emerges in the specific heat accompanied by a magnetization jump, indicating  phase transition to a long-range ordered state. 
\begin{figure}[ph]
\includegraphics[width=0.9\columnwidth,clip]{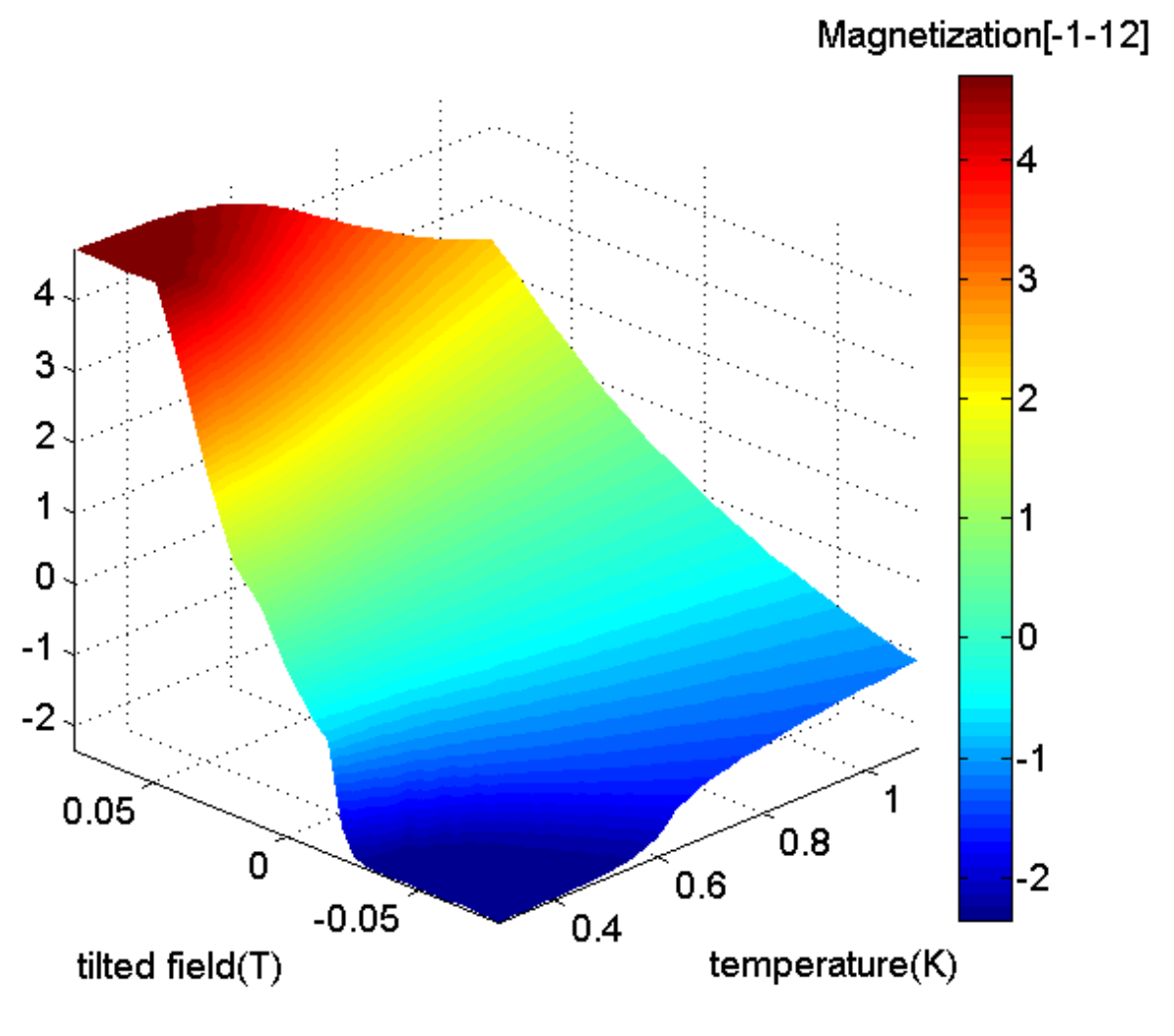}
\caption{(Color online) Magnetization  along $[\bar{1} \bar{1} 2]$ direction as a function of tilted fields and temperature. A clear jump is observed  at low temperatures for negative tilted field.}
\label{fig:mag}
\end{figure}

\section{Magnetization in the kagome plane}
Figure~\ref{fig:mag} shows the  magnetization in the kagome plane in  positive and negative field tilts. 
The dark red region refers to the fully-polarized state  with  saturated value of magnetization per spin  in the  4.714 $\mu_{B}$  for a large tilted field, consistent with that of a fully-polarized state, as found in a [100] field~\cite{jch+2008,Lin:2013kh}.  The deep blue region refers to the $\mathbf{q} = X$ state. 
A magnetization jump is associated with the peak of the specific heat for negative tilted field $|H_\perp| > 0.03$~T, indicating the transition to the $\mathbf{q}=X$ state is first-order.

\end{document}